\documentclass{article}

\usepackage[final]{neurips_2023}

\usepackage[utf8]{inputenc} 
\usepackage[T1]{fontenc}    
\usepackage{hyperref}       
\usepackage{url}            
\usepackage{booktabs}       
\usepackage{amsfonts}       
\usepackage{float}          
\usepackage{nicefrac}       
\usepackage{microtype}      
\usepackage{xcolor}         
\usepackage{graphicx}
\usepackage{amsmath}
\usepackage{subcaption}
\usepackage{mathtools}
\usepackage{wrapfig}

\title{Inferring Dynamic Hidden Graph Structure in Heterogeneous Correlated Time Series}

\author{%
  Jeshwanth Mohan \\
  Dept of Electrical Engineering \& Computer Science \\
  University of California, Berkeley \\
  \texttt{jeshum29@berkeley.edu}
  \And
  Bharath Ramsundar \\
  Deep Forest Sciences \\
  \texttt{bharath@deepforestsci.com}
  \And
  Sandya Subramanian \\
  Dept of Computational Precision Health \\
  University of California, Berkeley \\
  \texttt{sandyas@berkeley.edu}
}

\begin{document}
\maketitle
\vspace{-0.25in}
\begin{abstract}
 Modeling heterogeneous correlated time series requires the ability to learn hidden dynamic relationships between component time series with possibly varying periodicities and generative processes. To address this challenge, we formulate and evaluate a windowed variance-correlation metric (WVC) designed to quantify time-varying correlations between signals. This method directly recovers hidden relationships in an specified time interval as a weighted adjacency matrix, consequently inferring hidden dynamic graph structure. On simulated data, our method captures correlations that other methods miss. The proposed method expands the ability to learn dynamic graph structure between significantly different signals within a single cohesive dynamical graph model.
\end{abstract}

\vspace{-0.2 in}

\section{Introduction}

Modeling time-varying interacting phenomena as a dynamical network is a popular approach across many fields like neuroscience and pandemic modeling. Different phenomena are represented as nodes and their interactions as time-varying edges [1]. A significant technical challenge arises when modeling systems where different processes have varying generative models, such as in the case of different organ systems or different industries in an economic model. Traditional time series methods often assume homogeneity across node types, and are not designed for this type of multimodal data and can fail to produce meaningful results [2]. To overcome this, we define an edge metric designed to quantify correlations between heterogeneous time series. 

\section{Methods}

Suppose we have a $d$-dimensional time series signal $X \in \mathbb{R}^{d \times L}$ where $X_t \in \mathbb{R}^d$ has $d$ different components $X_{t,i}$. We seek to infer undirected edges $u_{t;i,j}$ between these different components to determine a graph structure $G_t = \{X_{t,i}, u_{t;i,j}\}$ for time $t$. The graph structure in the data is dynamic, so $G_t$ may vary over time.

Each signal $X_{t,i}$ may have a differing periodicity. We consequently define signal-specific window lengths $\tau_i$ based on the autocorrelation. We choose windows with fractional overlap $\alpha$. Setting $\beta = \left\lfloor\frac{L}{\tau(1-\alpha)}\right\rfloor$ to be the number of windows, we compute the means and standard deviations for each position within a window as,
\[
\mu_{X,i}(k) = \frac{1}{\beta_i} \sum\limits_{u=0}^{\beta_i} X_i(k + u \tau_i(1 - \alpha))
\quad \text{ and } \quad
\sigma_{X,i}(k) = \sqrt{\cramped{\frac{1}{\beta_i} \sum\limits_{u=0}^{\beta_i} X_i(k + u \tau_i(1 - \alpha))^2 - \mu_{X,i}(k)^2}}
\]
where $k$ ranges in $\{1, \dotsc, \tau_i\}$ and we follow the notation $X_i(t) = X_{t,i}$ for readability. Thus $\mu_{X,i} \in \mathbb{R}^{\tau_i}, \sigma_{X,i} \in \mathbb{R}^{\tau_i}$ are window-length vectors. We normalize the signals with respect to periodic behavior as,

$$Z_{X,i}(t) = \frac{X_i(t) - \mu_{X,i}(t \text{ (mod } \tau_i))}{\sigma_{X,i}(t \text{ (mod } \tau_i))}.$$

We then define the correlation metric between signals $X_i$ and $X_j$ in the time interval $1 < t_1 < t_2 < L$ as,
\[\textrm{WVC}^{t_1,t_2}_{i,j} = \sum\limits_{t=t_1}^{t_2} Z_{X,i}(t)Z_{X,j}(t).\]

We directly use the computed correlation metric as a weighted adjacency metric, so $\textrm{WVC}^{t_1, t_2}_{i,j}$ specifies weighted graph $G^{t_1, t_2}$ for the interval $[t_1, t_2]$. We can choose multiple choices of $t_1, t_2$ to recompute graph structures on demand.

\subsection{Data \& Evaluation}
We use simulated data to test WVC. We simulate two independent time series: (1) a sine wave with a period of 150 seconds and (2) a repeating inverse Gaussian with a period of 240 seconds. The simulated data was constructed with differing periods and underlying functions to assess the robustness of our proposed method. We further test our method by modulating segments within the synthetic data; specifically, we multiply data points within corresponding segments by 1.1 or 0.9 to mimic local elevation or depression of the signal.

We seek a method to compare different choices of metric for graph recovery. Suppose we have two metrics $C_1$ and $C_2$. Suppose we can compute compute the Z-score of metrics $C_1, C_2$ over time. For uncorrelated signals, this Z-score should be close to $0$, and for correlated signals, the Z-score should be nonzero. 

We utilize the Fisher transformation to yield a normally distributed metric for the PCC, from which we compute Z-scores. Under a simplifying assumption that time series $X_i$ and $X_j$ are independent, we can derive that $\textrm{Var}(\textrm{WVC}_{i,j}) = (t_2 - t_1 + 1)\beta_i \beta_j$.


Using this variance, we can compute Z-scores for WVC. 
We can use the Z-scores to compute a probability for whether the signals are uncorrelated or correlated. We can then compute a RMS for this probability against the ground truth of uncorrelated or correlated.
\section{Results and Discussion}
\label{headings}

Fig. 1 shows WVC's performance compared to the PCC on simulated data, for independent signals and for signals where a relationship was artificially modulated. With the independent signals (Fig. 1, left panels), which are designed to have different periods and no true underlying relationship, our method correctly reports a near-zero correlation unlike the strong correlations that are incorrectly detected by the PCC. Conversely, for the modulated signals where periods of positive and negative correlation were deliberately introduced (Fig. 1, right panels), our metric clearly and accurately captures these dynamic changes. 
\begin{figure}[H]
    \centering 

    \begin{subfigure}[b]{0.49\textwidth}
        \includegraphics[width=\textwidth]{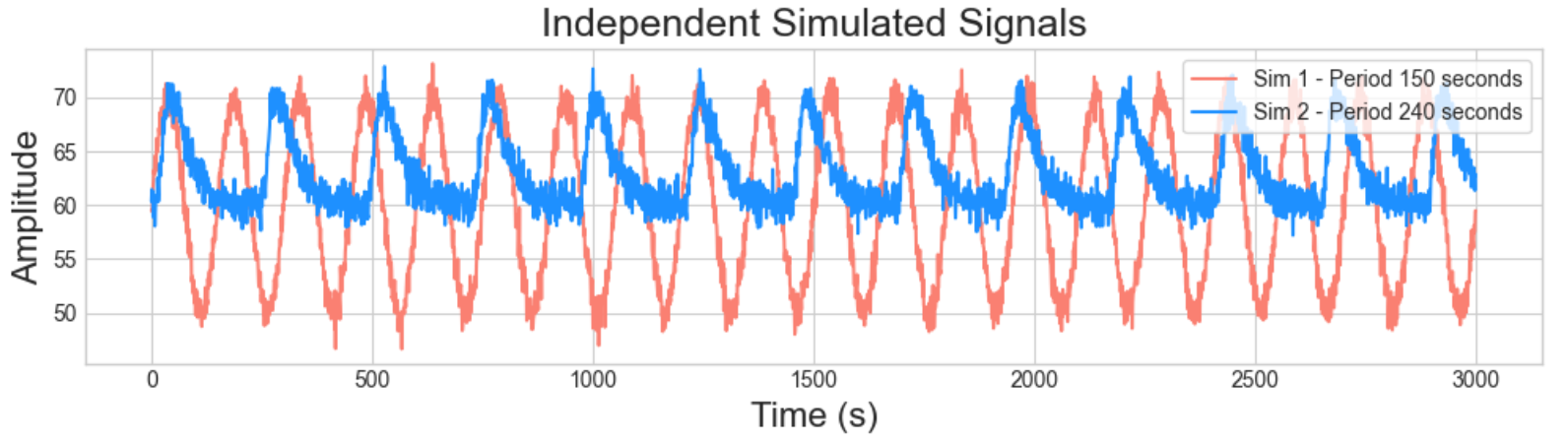}
    \end{subfigure}
    \begin{subfigure}[b]{0.49\textwidth}
        \includegraphics[width=\textwidth]{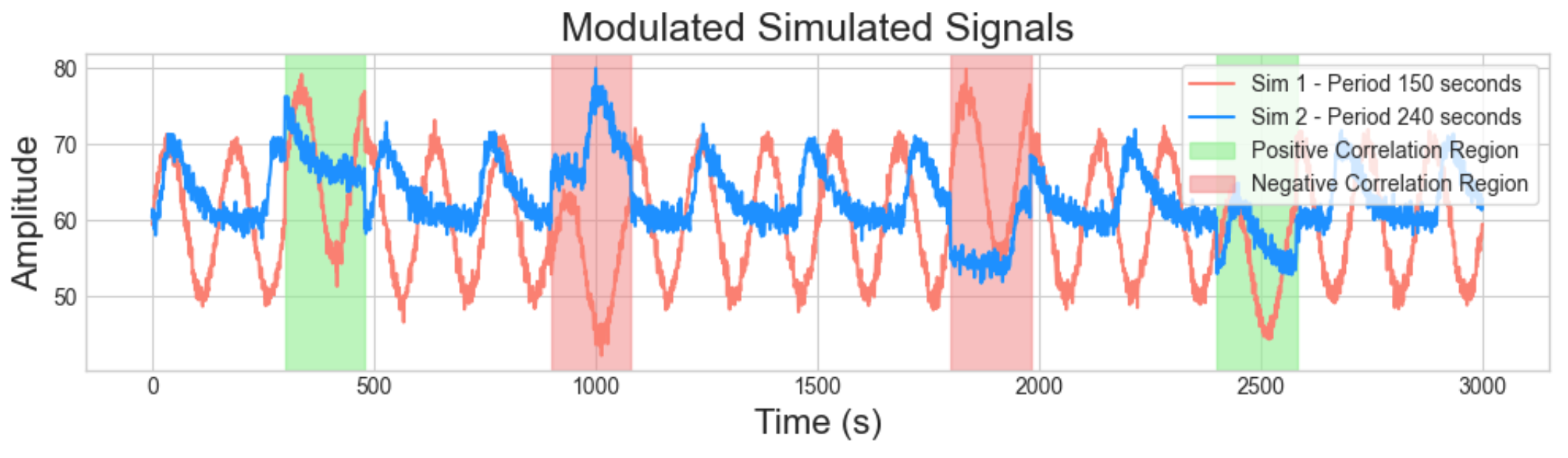}
    \end{subfigure}
    \begin{subfigure}[b]{0.49\textwidth}
        \includegraphics[width=\textwidth]{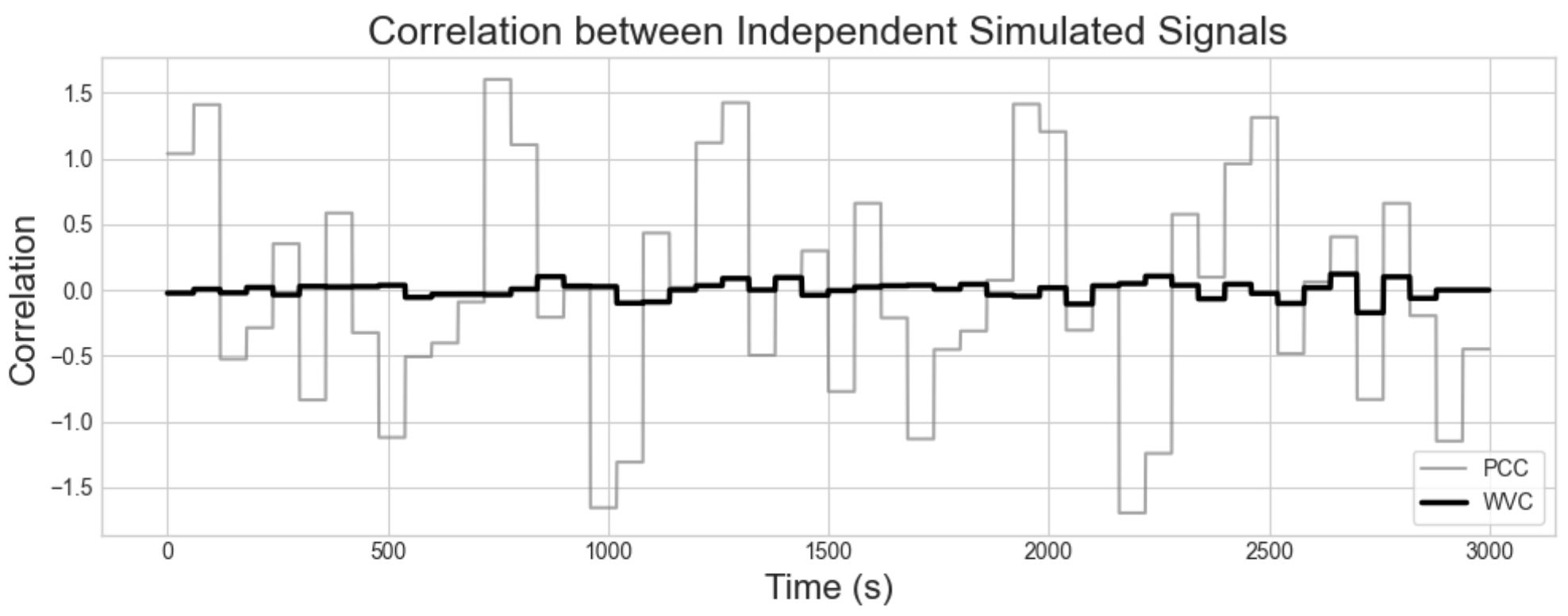}
    \end{subfigure}
    \begin{subfigure}[b]{0.49\textwidth}
        \includegraphics[width=\textwidth]{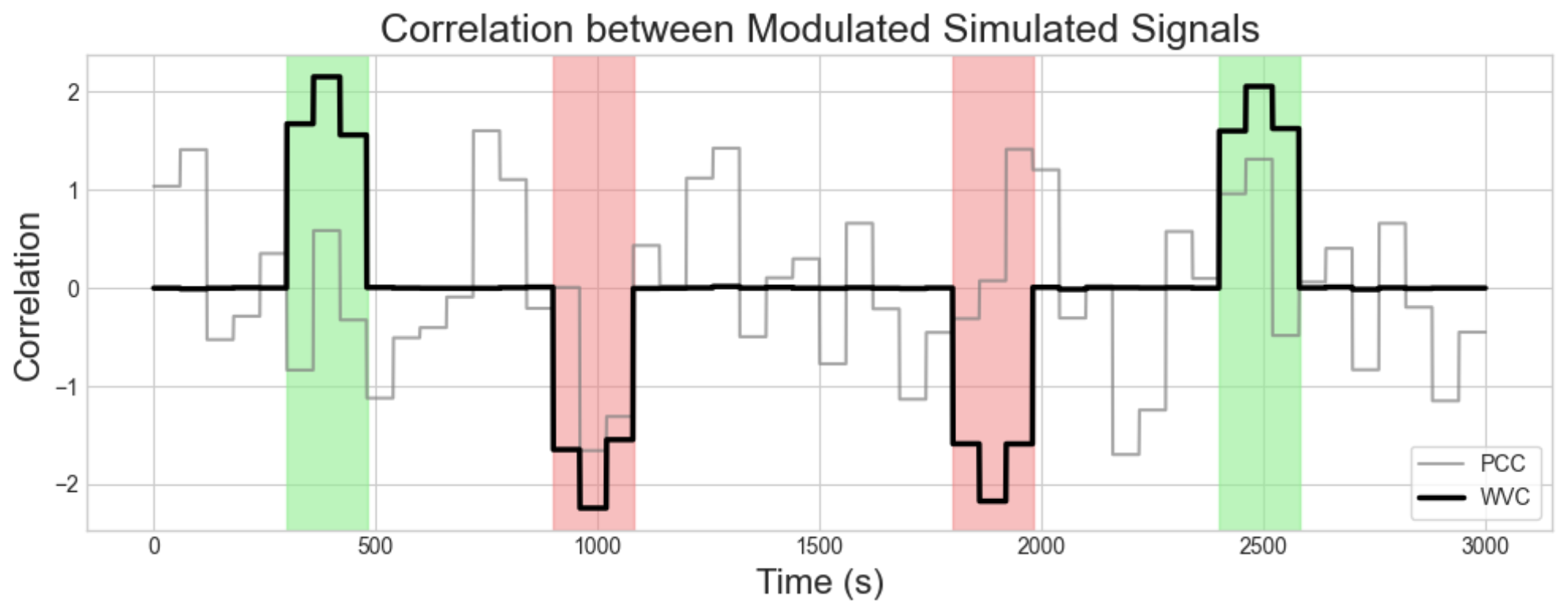}
    \end{subfigure}
    \small{
    \caption{Simulated versus modulated data and their respective correlation metrics.}}
    \label{fig:main_grid}
\end{figure}
    
\begin{wraptable}{r}{6.cm}
\vspace{-0.5cm}
\caption{RMSE of PCC and WVC to ground truth.}
\begin{tabular}{lcc}
      \toprule 
      & \textbf{Independent} & \textbf{Modulated} \\
      \midrule 
      \textbf{PCC} & 0.5343  & 0.5342 \\
      \textbf{WVC} & 0.0470 & 0.0448  \\
      \bottomrule 
\end{tabular}
\end{wraptable}
    



WVC successfully expands the capability of dynamic graph learning    to identify time-varying correlations between signals exhibiting different periodicities and generative models. This ability to define an edge between heterogeneous signals, such as point processes and coupled oscillators, represents a leap for graph learning by expanding its applicability to more complex, multimodal systems, such as the human body.
\\\\
\section*{References}

‌

\small

[1] D. Hevey, “Network analysis: a brief overview and tutorial,” Health Psychology and Behavioral Medicine, vol. 6, no. 1, pp. 301–328, Jan. 2018.

[2] I. A. Iwok and A. S. Okpe, “A Comparative Study between Univariate and Multivariate Linear Stationary Time Series Models,” American journal of mathematics and statistics, vol. 6, no. 5, pp. 203–212, Jan. 2016.

[3] Bartlett, M. S. "On the Theoretical Specification and Sampling Properties of Autocorrelated Time-Series,"Supplement to the Journal of the Royal Statistical Society. 1946.
‌
\newpage
\vspace{0.5 in}
\section*{Appendix}

\subsection*{I. Derivations for Variance}

\subsubsection*{Setup}

Let $X$ be a $d$-dimensional time series signal, s.t. $X \in \mathbb{R}^{d \times L}$ where $X_t \in \mathbb{R}^d$ has $d$ different components $X_{t,i}$. Each observation, $X_{t,i}$ or $X_i(t)$  is drawn independently from a Normal distribution denoted by $X_i(t) \sim \mathcal{N}(\mu_i, \sigma_i^2)$, where $\mu_i$ and $\sigma_i$ are the global mean and standard deviations of $X_i$. We assume that the time series $X_i$ and $X_j$ are mutually independent for $i \neq j$.

Let $\beta_i$ represent the number of windows for signal $X_i$. We define the normalized time series $Z_i(t)$ as,
\begin{equation}
    Z_i(t) = \left( \frac{X_i(t) - \mu_{X,i}}{\sigma_{X,i}} \right) =  \sqrt{\beta_i} \left( \frac{X_i(t) - \mu_i}{\sigma_i} \right). \label{eq:def_Z}
\end{equation}

\subsubsection*{Derivation}

We seek to derive the variance of WVC. Assuming observations are independent over time, the variance of the sum equals the sum of the variances,
\begin{align}
    \text{Var(WVC)} &= \text{Var}\left( \sum_{t=t_1}^{t_2} Z_i(t) Z_j(t) \right) \\
    &= \sum_{t=t_1}^{t_2} \text{Var}\big( Z_i(t) Z_j(t) \big). \label{eq:sum_var}
\end{align}
Using the property $\text{Var}(Y) = \mathbb{E}[Y^2] - (\mathbb{E}[Y])^2$, we evaluate the term inside the summation. Since $X_i$ and $X_j$ are independent, $Z_i$ and $Z_j$ are independent. Thus,
\begin{equation}
    \mathbb{E}[Z_i(t) Z_j(t)] = \mathbb{E}[Z_i(t)] \cdot \mathbb{E}[Z_j(t)] = 0.
\end{equation}
Consequently, the variance simplifies to the second moment,
\begin{align}
    \text{Var}(Z_i(t) Z_j(t)) &= \mathbb{E}\left[ Z_i(t)^2 Z_j(t)^2 \right] \\
    &= \mathbb{E}[Z_i(t)^2] \cdot \mathbb{E}[Z_j(t)^2] \\
    &= \frac{\beta_i}{\sigma_i^2}\mathbb{E}[(X_i(t) - \mu_i)^2] \cdot \frac{\beta_j}{\sigma_j^2}\mathbb{E}[(X_j(t) - \mu_j)^2] \\
    &= \frac{\beta_i}{\sigma_i^2} \text{Var}(X_i(t)) \cdot \frac{\beta_j}{\sigma_j^2} \text{Var}(X_j(t)) \\
    &= \beta_i \beta_j.
\end{align}
Substituting this result back into \eqref{eq:sum_var},
\begin{equation}
    \text{Var(WVC)} = \sum_{t=t_1}^{t_2} \beta_i \beta_j = (t_2 - t_1 + 1)\beta_i \beta_j.
\end{equation}
\\\\
\subsection*{II. Determination of the Signal-Specific Window Length $\tau_i$}

As discussed in Section 2, the Windowed Variance-Correlation (WVC) method requires a signal-specific window length $\tau_i$ for each time series component $X_i$ to account for heterogeneous periodicities. This length is defined by the fundamental period of the signal, automatically determined using the Autocorrelation Function (ACF). The process is executed independently for each signal $X_i \in \mathbb{R}^{L}$, where $L$ is the time series length.

\subsection*{Autocorrelation Function (ACF) Calculation}

The initial step is to compute the Autocovariance Function (ACVF), denoted $R_{X X}(\tau)$, which measures the correlation of a signal with a delayed copy of itself at lag $\tau$. This function is computed for all lags $\tau = 1, 2, \ldots, \tau_{\max}$.

\begin{enumerate}
    \item \textbf{Demeaning:} The signal $X_i$ is first demeaned by subtracting its global mean $\mu_i$.
    \item \textbf{Autocovariance ($R_{X X}(\tau)$):} The sample ACVF for lag $\tau$ is defined as:
    $$R_{X_i X_i}(\tau) = \frac{1}{L-\tau} \sum_{t=1}^{L-\tau} (X_i(t) - \mu_i)(X_i(t+\tau) - \mu_i)$$
    \item \textbf{Normalized Coefficient ($\rho_{\tau}$):} The normalized autocorrelation coefficient $\rho_{\tau}$ is then computed by dividing $R_{X_i X_i}(\tau)$ by the variance, which is also $R_{X_i X_i}(0)$:
    $$\rho_{\tau} = \frac{R_{X_i X_i}(\tau)}{R_{X_i X_i}(0)}$$
\end{enumerate}

\subsection*{Significance Threshold}

The normalized coefficient $\rho_{\tau}$ is used to compare the observed correlation to the statistical noise floor expected from a white noise process. This comparison ensures that the detected period is statistically significant and not merely noise.

For a large time series that is assumed to be white noise, the $95\%$ confidence interval (or noise floor) for the ACF is approximated by [3]:
$$\text{Threshold} \approx \frac{1.96}{\sqrt{L}}$$
Any ACF value $|\rho_{\tau}|$ falling below this magnitude is considered statistically insignificant.

\subsection*{Automatic Period Search}

The window length $\tau_i$ is set to the lag $\tau^*$ corresponding to the first statistically significant local maximum in the ACF, starting from $\tau=1$.

\begin{enumerate}
    \item \textbf{Search Criteria:} The algorithm searches for the smallest lag $\tau^* > 1$ that satisfies two conditions simultaneously:
    \begin{itemize}
        \item \textbf{Significance:} $|\rho_{\tau^*}| > \frac{1.96}{\sqrt{L}}$
        \item \textbf{Local Maximum:} $\rho_{\tau^*} > \rho_{\tau^*-1}$ and $\rho_{\tau^*} > \rho_{\tau^*+1}$
    \end{itemize}
    \item \textbf{Set $\tau_i$:} If such a $\tau^*$ is successfully identified, the window length is set as $\tau_i = \tau^*$, else set $\tau_i = 1$.
\end{enumerate}

Setting $\tau_i = 1$ is valid for aperiodic signals as this is effectively a normalization across the whole time series.

\end{document}